\documentclass[useAMS,usenatbib]{mn2e}

\title[How to Detect the Signatures of Self-Gravitating Circumstellar Discs with ALMA]{How to Detect the Signatures of Self-Gravitating Circumstellar Discs with the Atacama Large Millimetre/sub-millimetre Array}
\author[Dipierro et al.]{
G.~Dipierro$^{1,2}$\thanks{E-mail: giovanni.dipierro@unimi.it}, 
G.~Lodato$^1$,
L.~Testi$^{2,3,4}$ and 
I. de Gregorio Monsalvo$^{2,5}$\\
$^1$ Dipartimento di Fisica, Universit\`a  degli Studi di Milano, Via Celoria 16, 20133 Milano, Italy\\
$^2$ European Southern Observatory, Karl Schwarzschild str. 2, D-85748 Garching bei M\"unchen, Germany\\
$^3$ INAF-Osservatorio Astrofisico di Arcetri, Largo E. Fermi 5, I-50125 Firenze, Italy\\
$^4$ Excellence Cluster Universe, Boltzmann str. 2, D-85748 Garching bei M\"unchen, Germany\\
$^5$ Joint ALMA Observatory, Alonso de C\'ordova 3107, Vitacura 763-0355 Santiago, Chile}
\usepackage[T1]{fontenc}  
\usepackage{graphicx}
\usepackage{xcolor}
\usepackage{hyperref}
\hypersetup{colorlinks,%
citecolor=black,%
filecolor=black,%
linkcolor=black,%
urlcolor=black}
\usepackage{indentfirst}
\usepackage[bf, small]{caption}
\usepackage{subfigure}
\usepackage{aecompl}
\usepackage{xspace}
\usepackage{amsmath}
\usepackage{epsfig}
\usepackage{verbatim}
\usepackage{setspace}
\usepackage{amsfonts}
\usepackage{color}
\usepackage{url}
\usepackage{booktabs}
\usepackage{fancyhdr}
\usepackage{textcomp}
\usepackage{natbib}
\usepackage{multirow}
\usepackage{textcomp}
\newcommand {\apgt} {\ {\raise-.5ex\hbox{$\buildrel>\over\sim$}}\ }
\newcommand {\aplt} {\ {\raise-.5ex\hbox{$\buildrel<\over\sim$}}\ }

\begin{document}
\date{}
\maketitle
\begin{abstract}
In this paper we present simulated Atacama Large Millimetre/sub-millimetre Array (ALMA) observations of self-gravitating circumstellar discs with different properties in size, mass and inclination, located in four of the most extensively studied and surveyed star-forming regions.  Starting from a Smoothed Particle Hydrodynamics (SPH) simulation and representative dust opacities, we have initially constructed maps of the expected emission at sub-mm wavelengths of a large sample of discs with different properties. We have then simulated realistic observations of  discs as they may appear with ALMA using the Common Astronomy Software Application ALMA simulator. We find that, with a proper combination of antenna configuration and integration time, the spiral structure characteristic of self-gravitating discs is readily detectable by ALMA over a wide range of wavelengths at distances comparable to TW Hydrae ($\sim 50 \,$pc), Taurus - Auriga and Ophiucus ($\sim 140 \,$pc) star-forming regions.
However, for discs located in Orion complex ($\sim 400 \,$pc) only the largest discs in our sample (outer radius of 100 au) show a spatially resolved structure while the smaller ones (outer radius of 25 au) are characterized by a spiral structure that is not conclusively detectable with ALMA. 
\end{abstract}
\section{INTRODUCTION}

In the last fifteen years, it has been clearly recognized that the development of gravitational instabilities in protostellar discs plays a crucial role for their dynamical evolution. 
It is believed that in the early stage of star formation (Class 0/Class I objects), the disc could be massive enough to have a non-negligible dynamical effect on the evolution of the overall system. The disc self-gravity may affect the disc dynamics through the propagation of density waves which lead to the formation of a prominent spiral structure and provide a non-negligible contribution to the angular momentum transport (\citealt{LodatoRice2004,LodatoRice2005}).
In recent years, the investigation on the effects related to the disc self-gravity has been the subject of several theoretical studies aimed on the one hand at understanding angular momentum transport driven by gravitational instability (\citealt{Durisen}, \citealt{CossinsLodatoClarke2009}, \citealt{Forgan2011}) and, on the other hand, at exploring the possibility of planet formation either through disc fragmentation and by providing local pressure traps to promote grain growth and confinement (\citealt{MeruBate2011a,MeruBate2011b}, \citealt{Pinilla}, \citealt{Testi2014}). 
In addition, a number of evolutionary chemical models have been introduced in order to figure out how the spiral density waves excited by the gravitational instability influences the chemical evolution of both the gaseous and the solid component of the circumstellar discs (\citealt{Ilee2011}).
These models could be empirically confirmed through high angular-resolution observations. In particular, the detection of spiral features in circumstellar disc may allow to put constraints on the theoretical models about the physical and chemical evolution of the discs.

The macroscopic feature introduced by self-gravity is the creation of large-scale density fluctuations (in the form of a spiral pattern). To detect these features observationally and unambiguously identify them as density enhancement, it is necessary to observe at (sub-)mm wavelengths where the disc is mostly optically thin in the vertical direction and the emission from the mid-plane can be traced directly \citep{2007prpl.conf..555D}. This technique has been extensively used to probe the overall dust and molecular gas content since the deployment of the first sensitive observatories at millimetre wavelengths (\citealt{Beckwith90}, \citealt{1996A&A...309..493D}).  More recently, it has been possible to perform extensive surveys of discs in nearby star-forming regions that allowed us to constrain statistically their structural properties, molecular gas chemistry and dust evolution (\citealt{2011ARA&A..49...67W}, \citealt{2013ApJ...771..129A}, \citealt{2014arXiv1402.3503D}, \citealt{Testi2014}). While several studies have started to probe the physical structure and asymmetries of the outer discs  (\citealt{2005ApJ...622L.133C}, \citealt{2012ApJ...760L..17P}, \citealt{2012A&A...547A..84T}, \citealt{2013A&A...558A..64T}), spatially resolved images of protoplanetary discs with sufficient sensitivity and angular resolution to probe structures on the au scale (the expected scale of the spiral structure) are still very limited. The recently inaugurated ALMA observatory \citep[see][]{2013Msngr.152....2T}  will transform this field by providing observations with unprecedented resolution and sensitivity that will allow us to resolve features in the inner regions of protoplanetary discs (\citealt{2007ApJ...665..478K}, \citealt{2008NewAR..52..105T}, \citealt{CossinsLodatoTesti2010}). Initial results from Science Verification and Early Science are already showing the capabilities of ALMA to study the small-scale physical and chemical structure of protoplanetary discs (\citealt{2013Natur.493..191C}, \citealt{2013A&A...557A.133D}, \citealt{2013A&A...557A.132M}, \citealt{2013Sci...340.1199V}). It is thus very timely to investigate how future ALMA observations could be optimized to study the expected effects of self-gravity in discs.
A first attempt at assessing, based on the expected properties of the gravitational instability, the detectability of a spiral structure in the disc has been made by \citet{CossinsLodatoTesti2010} who have found that spiral structures in compact self-gravitating discs are visible with ALMA over a wide range of wavelengths.
Moreover, direct detections of planet-induced structures using high angular resolution (sub-)millimetre wave interferometers such as ALMA has been obtained by \citet{WolfAngelo2005} and recently by \citet{Ruge2013}. 
In this context, high-resolution, direct imaging observations of circumstellar/protoplanetary discs have already revealed non-axisymmetric structures which are believed to be due to embedded planets or external tidal interactions (\citealt{GradyMuto2012}, \citealt{GradyMuto2013} and \citealt{Christiaens2014}).
Furthermore, using models of molecular line emission from planet-forming circumstellar discs, \citet{Narayanan2006} have shown that dense gas clumps associated with gas giant planet formed via disc instabilities may be observable with ALMA. 
Moreover, processes such as desorption of various chemical species from the surface of disc dust grains and gas-phase chemical reactions due to spiral shocks occurring in the spiral arms of gravitationally unstable discs, produce clear chemical signatures of the disc dynamics \citep{Ilee2011} which may be detected using ALMA with the aim to infer the physical structure of self-gravitating discs \citep{Douglas2013}. 

The theoretical characterization of gravitational instability conducted by \citet{CossinsLodatoClarke2009} has shown a number of interesting features which could be empirically confirmed by high-resolution observations. For example, features such as the opening angle and the size of spiral arms are related to the disc-to-star mass ratio. Thus, the detection of spiral arms may be useful to infer some physical characteristics about the star-disc system.
In addition, it is expected that the density inhomogeneities induced by the development of gravitational instabilities affects both the radial migration and the growth of the dust grains (\citealt{RiceLodato2004,RiceLodato2006}, \citealt{Pinilla}).
In this paper we demonstrate whether the spiral structure excited by the gravitational instability can be detected using ALMA. 
This work extends the previous analysis of \citet{CossinsLodatoTesti2010} through the evaluation of ALMA simulated observations of a wide set of both face-on and non face-on discs with different properties in size and mass and a more accurate computation of the dust opacity. 

The paper is organized as follows: in section \ref{sec:numeri_setup} we describe the simulation details and the disc evolution. In section \ref{sec:generation}, we discuss the steps in order to create intensity maps and in section \ref{sec:alma_simul} we outline the basic steps for an ALMA simulated observation. Finally, in section \ref{sec:results} we present some of the ALMA images of our sample and in section \ref{sec:discuss} discuss the significance of our results.

\section{DISC MODEL}
\label{sec:numeri_setup}
\begin{figure*}
\begin{minipage}{\textwidth}
\centering
\includegraphics[scale=0.32]{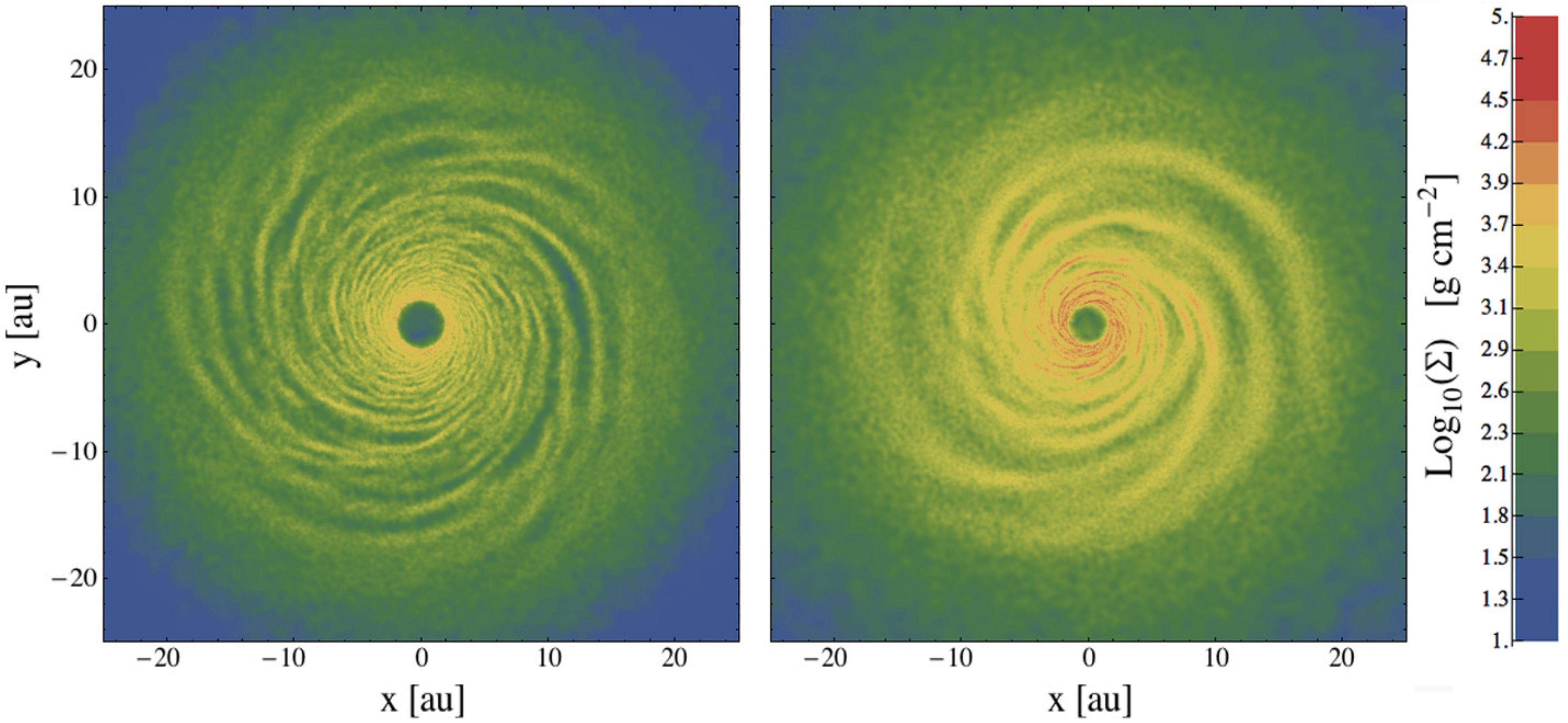}
\caption{Surface density structure at the end of the simulations with radius 25 au and $M_{\star}=1\,\rm{M_{\odot}}$ and two different disc-to-star mass ratio: (left) $q=0.1$; (right) $q=0.25$.}
\label{img:plot_density}
\end{minipage}
\end{figure*}
The simulations used to generate the expected disc emission have been taken from \cite{LodatoRice2004}. They performed simulations using a 3D smoothed particle hydrodynamics (SPH) code, a Lagrangian
hydrodynamics code (see reviews of \citealt{Monagnan2005} and \citealt{Price2010}). The system comprises a central star modelled as a point mass $M_{\star}$  surrounded by a gaseous disc of mass $M_{\rm{disc}}$ which extends from $R_{\rm{in}}=0.25$ to $R_{\rm{out}}=25$ in code units.
The simulations were performed using two values for the disc-to-central object mass ratio $q=M_{\rm{disc}}/M_{\star}=0.1$ and $q=0.25$. 
Regarding the thermal aspect of disc evolution, the heating is governed by both \textit{PdV} work and viscous dissipation while the cooling is implemented using a simple cooling law given by:
\begin{equation}
\frac{d u_{\rm{i}}}{dt}=-\frac{u_{\rm{i}}}{t_{\rm{cool},\rm{i}}}\, ,
\label{tcool}
\end{equation}
where $u_{\rm{i}}$ is the specific internal energy and $t_{\rm{cool},i}$ is the cooling time associated with the $i^{th}$ particle. The latter is determined using the simple parametrization $\Omega_{\rm{i}} \, t_{\rm{cool},\rm{i}}=\beta_{\rm{cool}}$, where $\beta_{\rm{cool}}$ is held constant and equal to $7.5$.  
The choice of $\beta_{\rm{cool}}$ value is critical because the cooling modifies the dynamical evolution of the disc leading to the formation of bound objects in sufficiently high-resolution simulations (\citealt{MeruBate2011a}, \citealt{LodatoClarke2011}). However, our analysis is aimed at investigating the observability of the ``grand design'' spiral structure of a self-gravitating disc and we do not consider phenomena occurring at small scales. We have thus used a value of $\beta_{\rm{cool}}$ that, at the numerical resolution used in \citet{CossinsLodatoClarke2009}, does not lead to fragmentation. 

In order to consider discs with realistic different properties in mass and radial extension, we modify the length and the mass scale of the data output of the SPH simulations. In this way, we can produce discs with properties consistent with the expected typical values obtained in observational campaigns (see the review of \citealt{2011ARA&A..49...67W}). We therefore consider two different values of the code unit length such that the radial extensions of the disc are $25$ au or $100$ au. Regarding the mass of the disc-star system, we use three different value of the mass scale such that the central mass star is equal to $0.3\,\rm{M_{\odot}}$, $1\, \rm{M_{\odot}}$ and $3\,\rm{M_{\odot}}$ and with disc mass covering a range from $0.075\, \rm{M_{\odot}}$ to $0.75\,\rm{M_{\odot}}$.

\subsection{Disc evolution}
\label{disc:evol}
Once the thermal equilibrium state is reached, the disc is characterized by a spiral structure induced by self-gravity. 
The spiral features are clearly seen in Fig. \ref{img:plot_density} where the (logarithmic) surface density is shown for the two simulations with radius 25 au and $M_{\star}=1\,\rm{M_{\odot}}$ and $q=0.1$ (left) and $q=0.25$ (right). It can be noted that, as the disc-to-star mass ratio $q$ increases, the pattern of the instability becomes more dominated by low number of arms and, in addition, the spiral structures tends to become more open. 
Also, as the disc mass increases, a marginally stable disc becomes thicker (since $H/R\sim q$) and the typical size of the spiral arms, which is of the order of $H$, thus increases. A detailed Fourier analysis of the expected morphology of the spiral can be found in \citet{LodatoRice2004} and \citet{CossinsLodatoClarke2009}.

\begin{figure}
\centering
\includegraphics[scale=0.44]{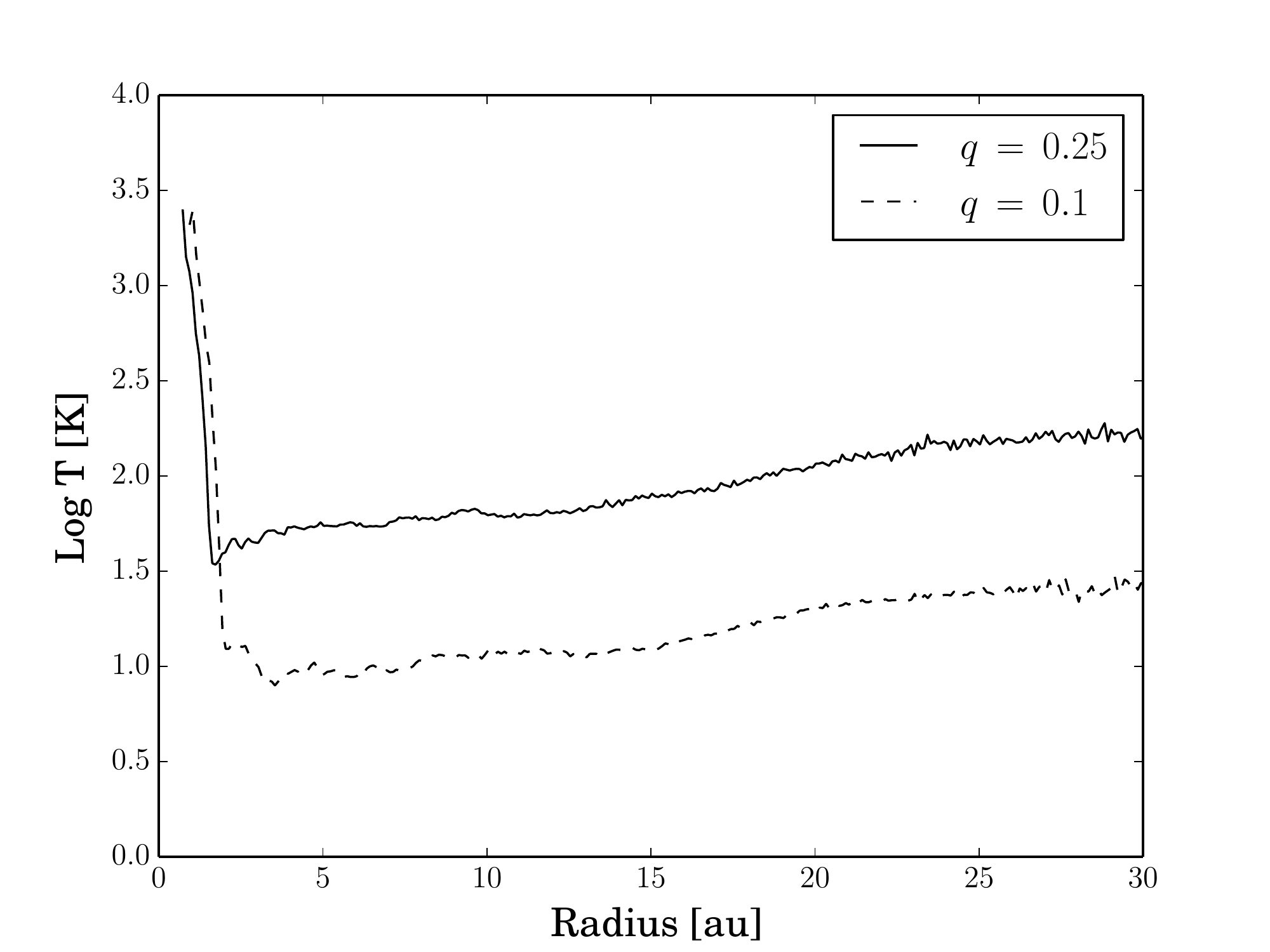}
\caption{Azimuthally averaged temperature profiles of the two discs with radius 25 au, $M_{\star}=1\,\rm{M_{\odot}}$ and $q=0.25$ (solid line) and $q=0.1$ (dashed line) once they are settled into the dynamic thermal equilibrium state.}
\label{img:temp_prof}
\end{figure}

In Fig. \ref{img:temp_prof} we show the azimuthally averaged temperature profiles associated with the two discs under consideration. The radial trend and the difference between profiles are linked to the thermal equilibrium state reached in the discs in which the cooling is balanced by heating generated through gravitational instabilities. 

\begin{figure}
\centering
\includegraphics[scale=0.42]{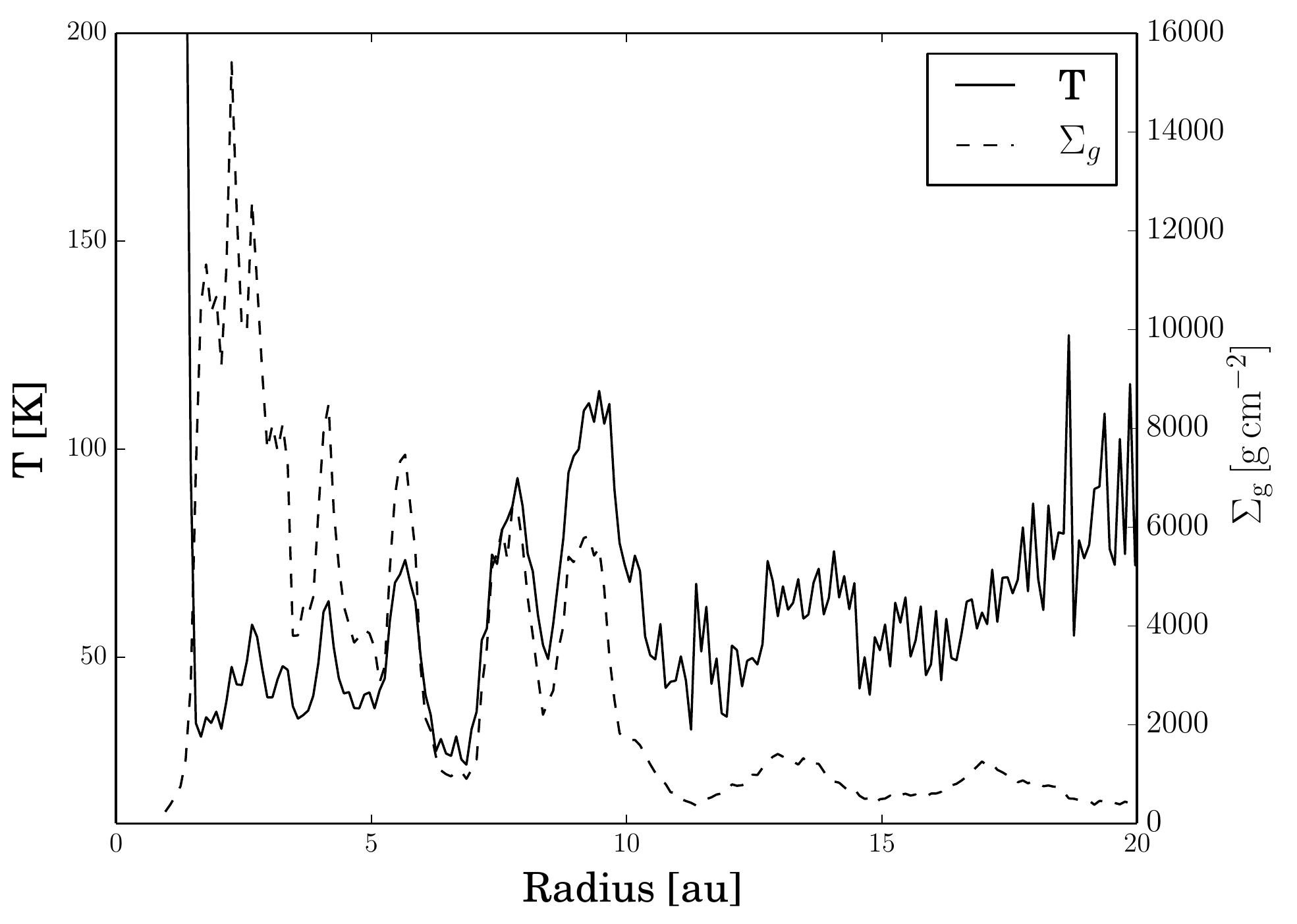}
\caption{Temperature (solid line) and density (dashed line) profiles for the disc with radius 25 au, $M_{\star}=1\,\rm{M_{\odot}}$ and $q=0.25$. The x-axis indicates the distance from the star along a line across the disc.}
\label{img:prof_temp_dens}
\end{figure}

To highlight the effect of the spiral on the gas and temperature structure within the disc, we show in  Fig. \ref{img:prof_temp_dens} the temperature and density profiles related to the disc with radius 25 au, $M_{\star}=1\,\rm{M_{\odot}}$ and $q=0.25$ along a horizontal slice across the disc. As could have been expected, the peaks in temperature occur at the same positions of the peaks in density showing that the overdensity regions, which correspond to the arm regions, are characterized by relatively higher temperatures. This is due to gas compression in the spiral arms.
From an observational point of view, the density and temperature bumps occurring in spiral arms allow the peculiar spiral structure of self-gravitating discs to be more easily detectable.

Note that the self-regulation of the gravitational instability, which is established by the thermal saturation of the instability, results in a slightly increasing azimuthally averaged temperature profile (see \ref{img:temp_prof}), which is not expected for a typical passive irradiated disc around pre-main-sequence stars. In fact, the disc model adopted here does not take into account the irradiation from the central star. Assessing the influence of this additional heating in the evolution of a self-gravitating disc is beyond the scope of this paper.  
In general, we expect that such extra heating source would weaken the amplitude of the spiral perturbations somewhat  (see \citealt{CossinsLodatoClarke2009} and \citealt{Riceiraddiation}). Thus, from an observational point of view, the detection of the spiral structure would probably be more challenging due to the decrease of the arm-interarm density contrast.

\section{GENERATION OF INTENSITY MAPS}
\label{sec:generation}
Starting from the output of the SPH simulations described above, it is then necessary to use the disc parameters to create intensity maps of the objects previously simulated. 
Following the approach of \citet{CossinsLodatoTesti2010}, we assume that the disc is in thermal equilibrium at temperature $T$ and is vertically isothermal. 
Therefore the source function at frequency $\nu$ is given by:
\begin{equation}
B_{\nu}(T)=\frac{2 h \nu^{3}}{c^{2}} \frac{1}{e^{h\nu/k_{B}T}-1}\, ,
\end{equation}
where $h$ is the Planck constant, $c$ is the speed of light and $k_B$ is Boltzmann's constant.
The temperature $T$ is obtained by vertically averaging the temperatures of all the particles inside a cylindrical radius equal to the smoothing length $h$ associated with the particle under consideration. 

At sub-mm wavelengths, dust emission is generally optically thin, meaning that it is necessary to estimate the optical depth in order to create intensity maps.
The optical depth $\tau_{\nu}$ at frequency $\nu$ is defined by:
\begin{equation}
\tau_{\nu}=\int_{-\infty}^{\infty}\kappa_{\nu}(z)\rho(z)\,dz \, ,
\end{equation}
where $\kappa_{\nu}(z)$ is the opacity of the disc at frequency $\nu$. 
Assuming that $z$ is the vertical component of the disc coordinate system and that the opacity is invariant with $z$, we can express the optical depth, for a face-on disc, as: $\tau_{\nu}=\kappa_{\nu}\Sigma $,
where $\Sigma$ is the total column density estimated by projecting the particles on the mid-plane and using a standard SPH formalism in 2D. 
Finally, we calculate intensity maps of discs with different properties in terms of mass, size and inclination using the following expression:
\begin{equation}
I_{\nu}=B_{\nu} \,(1- \exp (-\tau_{\nu})) \, ,
\label{eq:Inu}
\end{equation}
where the optical depth is appropriately computed taking into account also the case of non face-on discs. 

\subsection{Dust opacity}
In the high-density environment of a circumstellar disc, microns dust grains are expected to grow via collisional agglomeration and gravitationally settle towards the mid-plane of the disc (\citealt{NattaTestiCalvetHenning2007}, \citealt{Testi2014}). The evidence that grains in discs are significantly larger than grains in the diffuse interstellar medium (ISM) has been obtained from a variety of observational techniques (see the review of \citealt{Testi2014}). In this scenario, the dust opacity is obviously influenced by grain processing and therefore its value depends on the grain sizes, chemical compositions and shapes that cannot be easily probed by observations. 

Often, the dust opacity is expressed as a simple power-law function of wavelength appropriate for dust grains with a size distribution of the form \citep{Draine2006}:
\begin{equation}
n(a)\propto a^{-q} \qquad \rm{for} \quad a_{\rm{min}}<a<a_{\rm{max}} \, ,
\end{equation}
where $a_{\rm{min}}$ and $a_{\rm{max}}$ represent the minimum and maximum size of the grains. 
Assuming that the dust grains are sphere-shaped particles in the limit $a \ll \lambda$ where $\lambda$ is the radiation wavelength, it is possible to analytically infer the dust opacity law in terms of a power-law function of the wavelength given by:
\begin{equation}
\kappa_{\lambda}=\kappa_{0}\biggl(\frac{\lambda}{\lambda_{0}}\biggr)^{-\beta} \, ,
\label{eq:opacity_dust}
\end{equation}
where $\kappa_{0}$ represents a fiducial dust opacity at wavelength $\lambda_{0}$. 
The parameter $\beta$ represents a very good indicator of the level of grain growth.
While the typical value for ISM grains is $\beta_{\rm{ism}}=1.7$, several observations (\citealt{Testi2001,Testi2003}, \citealt{Natta2004}, \citealt{AndrewsWilliams2007}, \citealt{RicciNattaTesti2010}, \citealt{Testi2014}) have found that $\beta_{\rm{disc}}<\beta_{\rm{ism}}$,  which is naturally interpreted in terms of grain growth (e.g. \citealt{Draine2006}). 
It is worth noting that the dust opacity law expressed in eq. \ref{eq:opacity_dust} is not quantitatively accurate in the case where the particle size becomes comparable with the radiation wavelength.   

Generally, through sub-millimetre observations and making some assumptions on the chemical composition, shape and size distribution of the dust grains, it is possible to put constraints on the level of grain growth. For an optically thin disc in the Rayleigh-Jeans limit, the sub-millimetre spectral energy distribution (SED) has a power-law behaviour with frequency expressed by $F_{\nu} \propto \kappa_{\nu} \,\nu^2\propto\nu^{\alpha}$ where $\alpha$ is the spectral index. Thus, from eq. \ref{eq:opacity_dust}, the indices $\beta$ and $\alpha$ are related by $\beta =\alpha -2$. 
However, taking into account the possibility that the emission is not completely thin and not in the Rayleigh-Jeans limit, the simple fit of the disc SEDs by a power-law in frequency does not provide a unique interpretation about the level of grain growth (\citealt{Beckwith90}, \citealt{Mohanty2013}). In order to solve this problem, combining the determination of the sub-millimetre/millimetre SED with information on the disc extension from high-angular resolution interferometric observations, the $\beta$ value is derived fitting the data with appropriate disc models (see \citealt{Testi2014} for details).  
\begin{figure}
\centering
\includegraphics[scale=0.44]{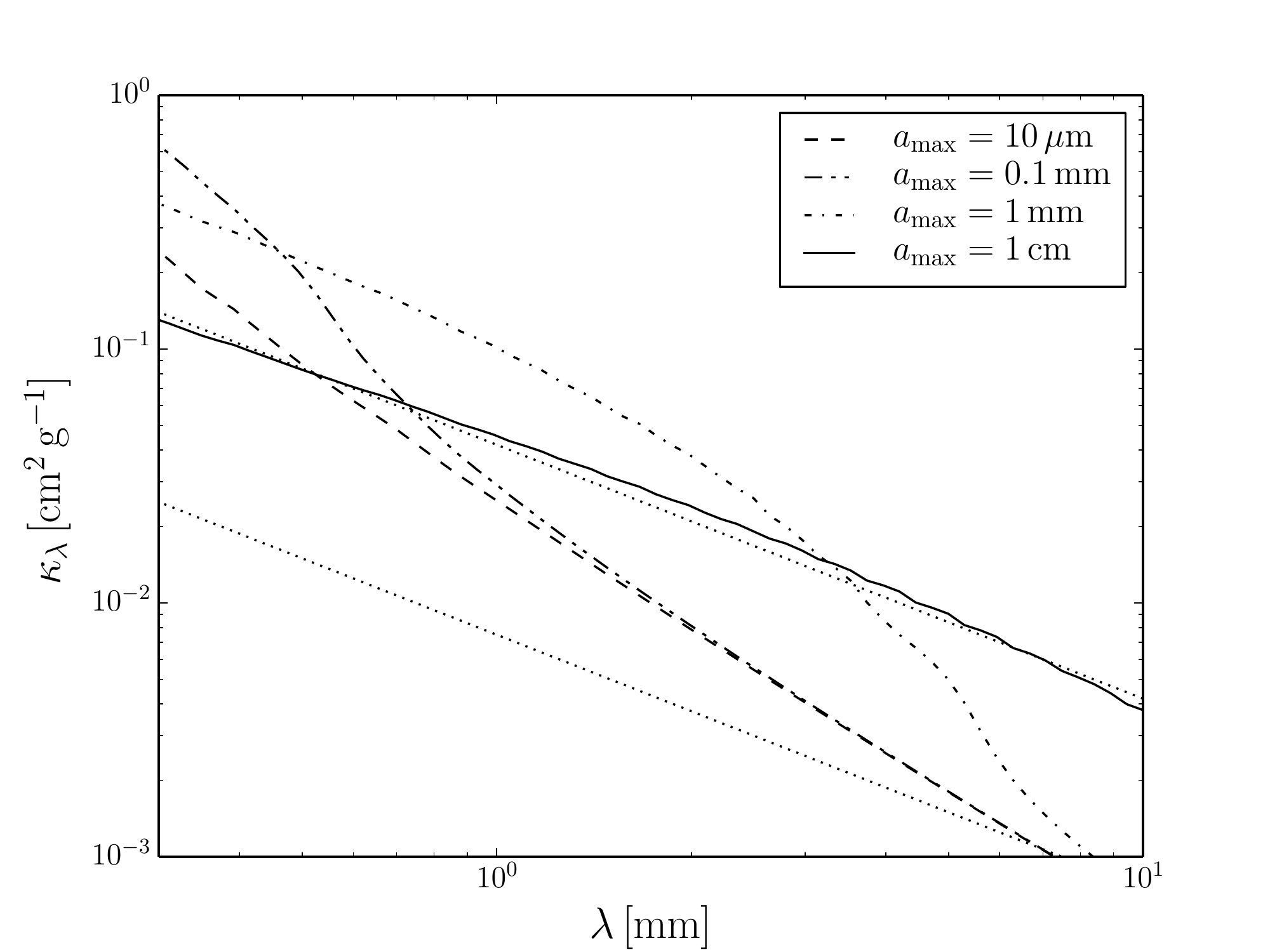}
\caption{Dust opacity for $a_{\rm{max}}$ between $10\, \rm{\mu m} -1 \,\rm{cm}$. Note that the power-law assumption, $k_{\lambda}\propto \lambda^{-\beta}$ breaks down for $a_{\rm{max}} \apgt \lambda$. The upper dotted line plots a power-law fit to the opacity data adopted here (solid line) and the lower dotted line represents the opacity values used by \citet{CossinsLodatoTesti2010}.}
\label{fig:opacit}
\end{figure}
In the light of the previous discussion, we compute the dust opacity using a particular dust model \citep{RicciNattaTesti2010} with an appropriate choice of model parameters. 
In Fig. \ref{fig:opacit}, the dust opacity values in units of $\rm{cm^2} \rm{g^{-1}}$ of gas are shown for the range of wavelengths under consideration computed for a grain size distribution $n(a) \propto a^{-3.5}$ between $a_{\rm{min}}=0.1 \,\rm{\mu m}$ and different values of $a_{\rm{max}}$. 
We use opacity values computed with $a_{\rm{max}} = 1\,\rm{cm}$ (solid line) for all the disc in order to take into account the grain growth processing. 
It can be noted that the dust opacity values adopted here approximately follow a power law (upper dotted line in Fig. \ref{fig:opacit}). 
In this respect, it is instructive to calculate the implied values of $\kappa_{\rm{1mm}}$ and $\beta_{\rm{1-3mm}}$ in this case: $\kappa_{1\rm{mm}} \approx 0.04$ $\mathrm{cm^2 g^{-1}}$ and $\beta_{\rm{1-3mm}} \approx 1$. The value of $\beta$ indicates that the dust opacities we use here take into account the grain growth.
Note that \citet{CossinsLodatoTesti2010} assume $\kappa_{\lambda} \approx 0.0075 \,(\lambda / \rm{1 mm})^{-1}\, \mathrm{cm^2 g^{-1}}$  (lower dotted line in Fig. \ref{fig:opacit}), a factor $\sim 5$ smaller than ours.

\section{THE ALMA SIMULATOR}
\label{sec:alma_simul}
We use the ALMA simulator in the Common Astronomy Software Application (CASA) package. 
We used the version 4.1 of CASA, which includes  simulation capabilities developed to simulate realistic ALMA observations over a wide range of wavelengths that would be measured by one of the full set of antenna configurations of ALMA. The simulation tool comes with 28 ALMA array configuration files containing the position of every antenna for each configuration.
In addition, it is also possible to add the expected receiver noise and the contribution due to the atmosphere that affect the visibility measurements. Using the Atmospheric Transmission at Microwaves (ATM) code \citep{Pardo}, the task \texttt{simobserve} calculates signal atmospheric corruption effects into the visibilities measurements by evaluating the atmospheric profile starting from average site conditions and input values for the ground temperature and the Precipitable Water Vapour (PWV) provided by the user.
Here, we assume a perfect calibration of the visibility measurements and thus only include the effects of thermal noise from receivers and the atmosphere.
This implies that for the highest angular resolution and highest frequency maps presented in this paper, the simulations may be optimistic. Nevertheless, at this stage, the experience with real ALMA observations on long baselines and high frequencies is extremely limited and it is difficult to assess how to properly account for these in the simulations.

Furthermore, having obtained the visibility measurements, the task \texttt{clean} operates a variety of algorithms in order to compute the deconvolved image from the visibility measurements. At the end of the process, it generates a cleaned image which represents the ALMA simulated observation.

\subsection{Image analysis}
The ALMA simulated observation can be analysed in terms of the ability to spatially resolve the spiral features of the circumstellar discs with an acceptable signal-to-noise ratio.

The noise of the cleaned image is obtained as the root mean square (rms) of the emission in regions where no signal is present, where we assume that the signal fluctuations are due to noise. Typically, it is used as a threshold for the signal detection and to calculate the signal-to-noise ratio. In this work, we use the typical minimum acceptable signal-to-noise ratio used for signal detection, i.e. equal to 3. 
It is worth remarking that each simulated observation has been performed with an appropriate choice of the antenna configuration and observation duration.  The aim is to ensure enough spatial resolution and sensitivity to resolve and detect spiral arms in the intermediate and external region of the discs with an acceptable signal-to-noise ratio. In particular, having evaluated the noise level in the resulting simulated observation images, we generate contour plots at different multiples of the evaluated rms noise which allow us to demonstrate that the spiral structure are detectable with an acceptable signal-to-noise ratio. 

\section{RESULTS}
\label{sec:results}
In this section, we present a sub-set of simulated ALMA observations of the discs previously described as they would be observed if they were located in four of the best-known and surveyed star-forming regions. 
In order to consider a wide range of distances and different sky positions, we locate the discs in the TW Hydrae ($\sim 50 \,$pc), Taurus - Auriga, Ophiucus ($\sim 140 \,$pc) and Orion ($\sim 400 \,$pc) star-forming regions. 
In this work, we use six observing frequencies chosen at the centre of some of the ALMA observation bands: 45 GHz in band 1, 100 GHz in band 3, 220 GHz in band 6, 345 GHz in band 7, 680 GHz in band 9 and 870 GHz in band 10. It is worth noting that, although the ALMA bands 1 and 10 are not yet available, we include these in our simulations as they are currently being deployed on the array (band 10) or in an advanced status of prototype development in view of full production (band 1). The array configuration used for each simulated observation varies for each frequency with the goal to use the configuration that offers the best compromise between resolution and sensitivity. 
For the atmospheric conditions of the observations, as discussed previously, we neglect the effect of signal decorrelation, assuming excellent calibration, but we include the effect of the atmospheric optical depth and emission parametrized through the amount of PWV. The values of the PWV that we use (see table \ref{tab:pwv}) are typical of the maximum matching constraints used at the observatory for the frequencies of our choice. 
\begin{table}
\begin{center}
\begin{tabular}{cc}
\midrule
Frequency (GHz)                 & PWV (mm) \\
\midrule
$45$             &     $2.748$\\
$100$          &     $2.748$\\
$220$          &    $2.748$\\
$345$          &    $1.262$\\
$680$          &    $0.472$\\
$870$          &    $0.472$\\
\bottomrule
\end{tabular}
\caption{PWV values for each observing frequency.}
\label{tab:pwv}
\end{center}
\end{table}
The choice of the PWV values does not play a crucial role at low frequencies because the atmospheric transmission is always very high at the centre of the band within the selected range of PWV. However, at higher frequencies, the PWV has to be carefully chosen due to the strong variation of the atmospheric noise with the PWV value. Thus, in real observation at higher frequencies ($> 200$ GHz), excellent observing conditions may be required for significant detection. We note that the values of the PWV assumed at 680 and 870 GHz are excellent, but better conditions are expected at Chajnantor for 15\% of the time. 
\subsection{Simulated ALMA images}
\begin{table*}
\setlength{\tabcolsep}{4.7pt}
\begin{minipage}{\textwidth}
\begin{center}
\begin{tabular}{ccccccccccccccccc}
\midrule
\midrule
\multicolumn{4}{c} {\footnotesize{Disc model parameters}} &  \multicolumn{6}{c} {\footnotesize{ALMA configurations}} & \footnotesize Duration&\multicolumn{6}{c} {\footnotesize{Total fluxes (mJy)}} \\
\midrule
\,\,\,\footnotesize{q}     &\, \footnotesize{$M_{\star} \,(\rm{M_{\odot}})$}     & \footnotesize{$R_{\rm{out}}$ (au)}  & \footnotesize{Incl. (\textdegree)}  &  \footnotesize45&\footnotesize100&\footnotesize220&\footnotesize345&\footnotesize680&\footnotesize870   &\footnotesize (min)    &  \footnotesize45&\footnotesize100&\footnotesize220&\footnotesize345&\footnotesize680&\footnotesize870 \\      
\,\,\,\footnotesize{(1)}     &\, \footnotesize{(2)}     & \footnotesize{(3)}  & \footnotesize{(4)}  &  \footnotesize{(5)}&\footnotesize{(6)}&\footnotesize{(7)}&\footnotesize{(8)}&\footnotesize{(9)}&\footnotesize{(10)}   &\footnotesize{(11)}    &  \footnotesize{(12)}&\footnotesize{(13)}&\footnotesize{(14)}&\footnotesize{(15)}&\footnotesize{(16)}&\footnotesize{(17)} \\  
\midrule
\multirow{3}{0.2cm}{\footnotesize0.25}& \multirow{3}{1.2cm}{\,\,\,\,\,\,\,\,\,\footnotesize0.3}& \multirow{3}{1.3cm}{\,\,\,\,\,\,\,\,\footnotesize25} & \footnotesize0  &\multirow{3}{0.1cm}{\footnotesize28} \,&\multirow{3}{0.1cm}{\footnotesize28}\,\,  & \multirow{3}{0.1cm}{\footnotesize25} \,\,  &\multirow{3}{0.1cm}{\footnotesize22} \,\, &\multirow{3}{0.1cm}{\footnotesize20}  \,\,&\multirow{3}{0.1cm}{\footnotesize18}\,\, &\multirow{3}{0.1cm}{\footnotesize30}\,\,& \footnotesize6 & \footnotesize37 &\footnotesize 158&\footnotesize 325 &\footnotesize 752 &\footnotesize 717 \\
& & & \footnotesize45  & & & &   & & && \footnotesize3 & \footnotesize20 &\footnotesize 88 &\footnotesize 172 &\footnotesize 420 &\footnotesize 403 \\
& & & \footnotesize65 & & & &   & & && \footnotesize1 & \footnotesize7 &\footnotesize 32 &\footnotesize 59 &\footnotesize 162 &\footnotesize 151 \\
\midrule
\multirow{3}{0.2cm}{\footnotesize0.25}& \multirow{3}{1.2cm}{\,\,\,\,\,\,\,\,\,\,\,\footnotesize1}& \multirow{3}{1.3cm}{\,\,\,\,\,\,\,\,\footnotesize25} & \footnotesize0 & \multirow{3}{0.1cm}{\footnotesize28}\,\, &\multirow{3}{0.1cm}{\footnotesize28}\,\,  & \multirow{3}{0.1cm}{\footnotesize25} \,\,  &\multirow{3}{0.1cm}{\footnotesize22} \,\, &\multirow{3}{0.1cm}{\footnotesize20} \,\, &\multirow{3}{0.1cm}{\footnotesize18}\,\, &\multirow{3}{0.1cm}{\footnotesize10}\,\,& \footnotesize26 & \footnotesize139 &\footnotesize 647 &\footnotesize 1466 &\footnotesize 4595 &\footnotesize 5837 \\
& & & \footnotesize45  & & & &   & & && \footnotesize13 & \footnotesize74 &\footnotesize 341&\footnotesize 788 &\footnotesize 2531 &\footnotesize 3357 \\
& & & \footnotesize65 & & & &   & & && \footnotesize4 & \footnotesize27 &\footnotesize 125 &\footnotesize 274 &\footnotesize 932  &\footnotesize 1244 \\
\midrule
\multirow{3}{0.2cm}{\footnotesize0.25}& \multirow{3}{1.2cm}{\,\,\,\,\,\,\,\,\,\,\,\footnotesize3}& \multirow{3}{1.3cm}{\,\,\,\,\,\,\,\,\footnotesize25} & \footnotesize0 &\multirow{3}{0.1cm}{\footnotesize28} \,&\multirow{3}{0.1cm}{\footnotesize28}  \,\, & \multirow{3}{0.1cm}{\footnotesize25} \,\,  &\multirow{3}{0.1cm}{\footnotesize22} \,\ &\multirow{3}{0.1cm}{\footnotesize20} \,\ &\multirow{3}{0.1cm}{\footnotesize18}\,\ &\multirow{3}{0.1cm}{\footnotesize10}\,\,& \footnotesize85 & \footnotesize454 &\footnotesize 2200&\footnotesize 4903 &\footnotesize 16186 &\footnotesize 21284 \\
& & & \footnotesize45  & & & &   & & && \footnotesize44 & \footnotesize239 &\footnotesize 1092&\footnotesize 2558 &\footnotesize 8890 &\footnotesize 13167 \\
& & & \footnotesize65 & & &   & & & && \footnotesize16 & \footnotesize85 &\footnotesize 398&\footnotesize 902 &\footnotesize 3276 &\footnotesize 4998 \\
\midrule
\,\multirow{2}{0.2cm}{\footnotesize0.1}& \multirow{2}{1.2cm}{\,\,\,\,\,\,\,\,\,\,\,\footnotesize3}& \multirow{2}{1.3cm}{\,\,\,\,\,\,\,\,\footnotesize25} & \footnotesize0 & \multirow{2}{0.1cm}{\footnotesize28}\,\,&\multirow{2}{0.1cm}{\footnotesize28}  \,\,& \multirow{2}{0.1cm}{\footnotesize27} \,\, &\multirow{2}{0.1cm}{\footnotesize24} \, &\multirow{2}{0.1cm}{\footnotesize22} \, &\multirow{2}{0.1cm}{\footnotesize20} &\multirow{2}{0.1cm}{\footnotesize120}\,\,\,\,& \footnotesize14 & \footnotesize71 &\footnotesize 364&\footnotesize 749 &\footnotesize 1232 &\footnotesize 2529 \\
& & & \footnotesize45  & & &  &  & & && \footnotesize7 & \footnotesize37 &\footnotesize 183&\footnotesize 346 &\footnotesize 704 &\footnotesize 1490 \\

\midrule
\multirow{3}{0.2cm}{\footnotesize0.25}& \multirow{3}{1.2cm}{\,\,\,\,\,\,\,\,\,\,\,\footnotesize1}& \multirow{3}{1.3cm}{\,\,\,\,\,\,\,\footnotesize100} & \footnotesize0 & \multirow{3}{0.1cm}{\footnotesize27}\,\, &\multirow{3}{0.1cm}{\footnotesize22} \,\, & \multirow{3}{0.1cm}{\footnotesize18} \,\,  &\multirow{3}{0.1cm}{\footnotesize15}\,\,\,  &\multirow{3}{0.1cm}{\footnotesize11} \,\, &\,\,\multirow{3}{0.1cm}{\footnotesize9} &\multirow{3}{0.1cm}{\footnotesize30}\,\,& \footnotesize50 & \footnotesize306 &\footnotesize 1324&\footnotesize 3173 &\footnotesize 5745 &\footnotesize 7317 \\
& & & \footnotesize45  & & &&    & & && \footnotesize28 & \footnotesize169 &\footnotesize 712&\footnotesize 1705 &\footnotesize 2993 &\footnotesize 4703 \\
& & & \footnotesize65 & & & &   & & && \footnotesize12 & \footnotesize68 &\footnotesize 282&\footnotesize 645&\footnotesize 1229 &\footnotesize 2045 \\
\midrule
\multirow{3}{0.2cm}{\footnotesize0.25}& \multirow{3}{1.2cm}{\,\,\,\,\,\,\,\,\,\,\,\footnotesize3}& \multirow{3}{1.3cm}{\,\,\,\,\,\,\,\footnotesize100} & \footnotesize0 & \multirow{3}{0.1cm}{\footnotesize27} \,&\multirow{3}{0.1cm}{\footnotesize22} \,\, & \multirow{3}{0.1cm}{\footnotesize18}\,\,\,\,&\multirow{3}{0.1cm}{\footnotesize15} \, &\multirow{3}{0.1cm}{\footnotesize11} \,\, &\,\,\multirow{3}{0.1cm}{\footnotesize9} &\multirow{3}{0.1cm}{\footnotesize10}\,\,& \footnotesize228 & \footnotesize1352 &\footnotesize 5553&\footnotesize 14257 &\footnotesize 24417 &\footnotesize 30786 \\
& & & \footnotesize45  & & &  &  & & && \footnotesize138 & \footnotesize655 &\footnotesize 2733&\footnotesize 7417 &\footnotesize 13751 &\footnotesize 17692 \\
& & & \footnotesize65 & & &   & & & && \footnotesize62 & \footnotesize277 &\footnotesize 1151&\footnotesize 2810 &\footnotesize 6913 &\footnotesize 9907 \\
\midrule
\midrule
\end{tabular}
\caption{Disc model, observation parameters and total resulting fluxes for the simulated observations of discs located in TW Hydrae star-forming region.
Column (1): disc-to-star mass ratio. Column (2): stellar mass. Column (3): outer radius. Column (4): disc inclination. Columns (5)-(10): ALMA configuration (see \url{http://casa.nrao.edu/} for details) for the six observing frequencies (expressed in GHz). Column (11): transit duration. Columns (12)-(17): total resulting flux at the various frequencies (expressed in GHz).}
\label{tab:twhya}
\end{center}
\end{minipage}
\end{table*}
In this subsection we present different simulations as a function of the frequency, disc-to-star mass ratio, stellar mass, inclination and coordinates of the discs. We analyse how the different values of the parameters mentioned above affects the detection of the spiral structures using ALMA. 
As discussed above, there are several physical and observational
parameters that determine the appearance of the discs with ALMA. We
have selected some specific configurations and parameter choices that
best show the possible outcome. The various parameters adopted in the
images presented here are summarized in \ref{tab:twhya}, \ref{tab:taurus} and \ref{tab:ophiucus}, referred to
discs located in TW Hydrae, Taurus-Auriga and Ophiucus star-forming region, respectively. In
particular, the first four columns specify the physical parameters of the disc (disc-to-star mass ratio, stellar mass, outer radius and inclination), columns (5)-(10) specify the ALMA configuration for the six bands that we have considered (where the number indicates the CASA identifier for the configuration of the various antennas, as described in \url{http://casa.nrao.edu/}), column (11) indicates the transit
duration for the simulated image, while columns (12)-(17) show the total resulting flux at the various frequencies (expressed in GHz).

\begin{figure*}
\begin{minipage}{\textwidth}
\centering
\includegraphics[scale=0.244]{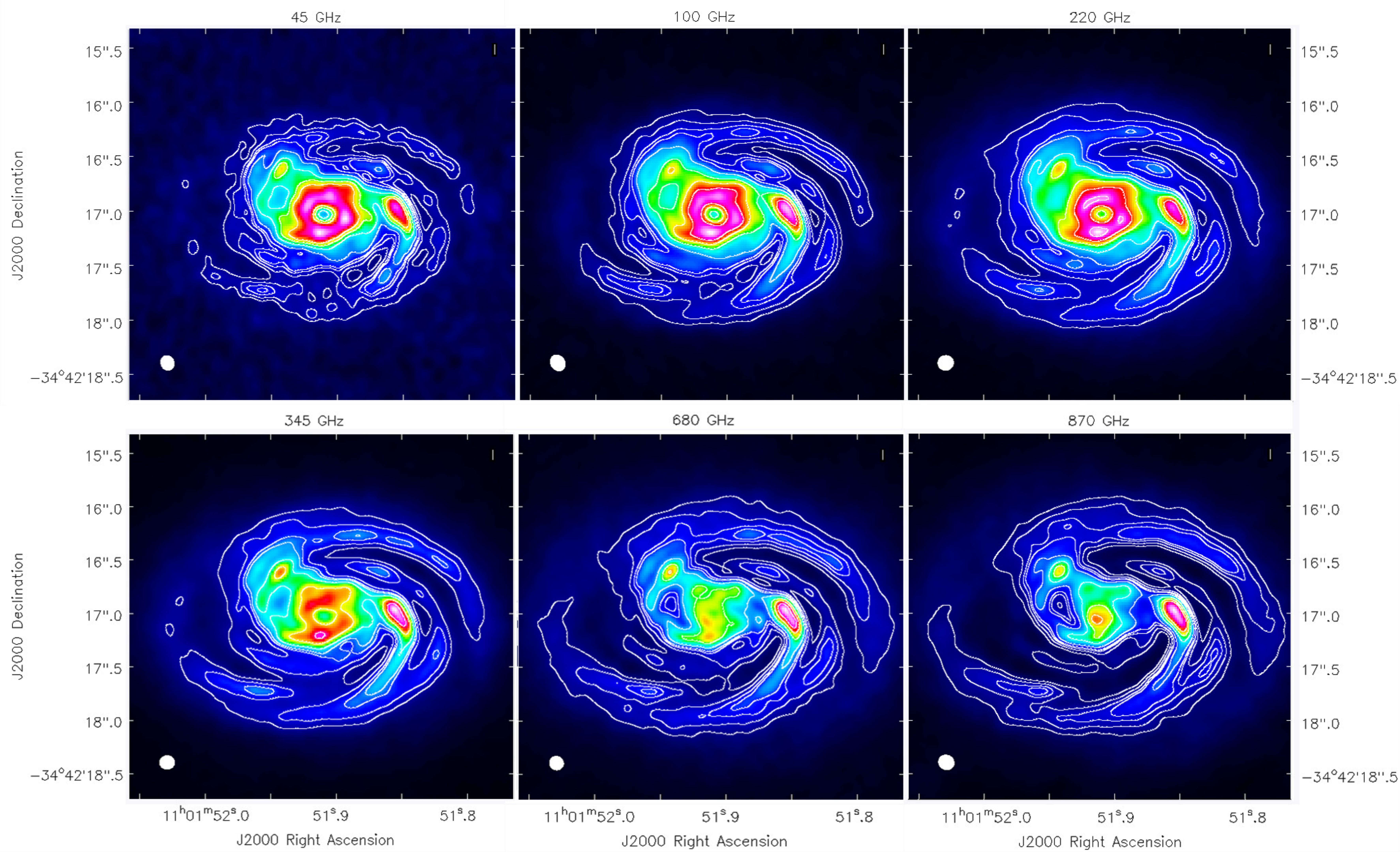}
\caption{ALMA images for the disc of radius 100 au with mass equal to $0.25\, \rm{M_{\odot}}$ and a disc-to-star mass ratio $q=0.25$ located in TW Hydrae. Contours are 7, 10, 13, 15, 20, 30, 40, 50 $\times$ the rms noise at 45 GHz, 100 and 220 GHz (7, 18, 27$\, \rm{\mu}$Jy/beam); 4, 7, 10, 13, 15, 20, 30, 40 $\times$ the rms noise at 345, 680 and 870 GHz (42, 260, 400$\, \rm{\mu}$Jy/beam). The white colour in the filled ellipse in the lower left corner indicates the size of the half-power contour of the synthesized beam: $6.6''\times6.1''$ (45 GHz), $7.2''\times6''$ (100 GHz), $7.3''\times6.7''$ (220 GHz), $7''\times6.2''$ (345 GHz),  $6.5''\times6.2''$ (680 GHz) and $7.3''\times6.7''$ (870 GHz).}
\label{img:alma_0.25_r4_m1_incl_45}
\end{minipage}
\end{figure*}

First, we show a representative set of ALMA images for all the observing frequencies of a disc with a given set of model parameters. In Fig.  \ref{img:alma_0.25_r4_m1_incl_45} we show the result of our simulations for non face-on discs (inclination angle of 45\textdegree) of radius 100 au located in TW Hydrae for a stellar mass of $1\, \rm{M_{\odot}}$ and disc-to-star mass ratio $q=0.25$ for the six selected observing frequencies (the original density structure of the disc is shown in Fig. \ref{img:plot_density}, right panel). In order to reach an optimal signal-to-noise ratio, the ALMA observations are simulated with a transit duration of 30 min. 
The resolution and sensitivity of ALMA are amply sufficient to spatially resolve and detect the substructures of the simulated discs over a wide range of wavelengths (confirming and extending the results of \citealt{CossinsLodatoTesti2010}). 

\begin{figure*}
\begin{minipage}{\textwidth}
\centering
\includegraphics[scale=0.25]{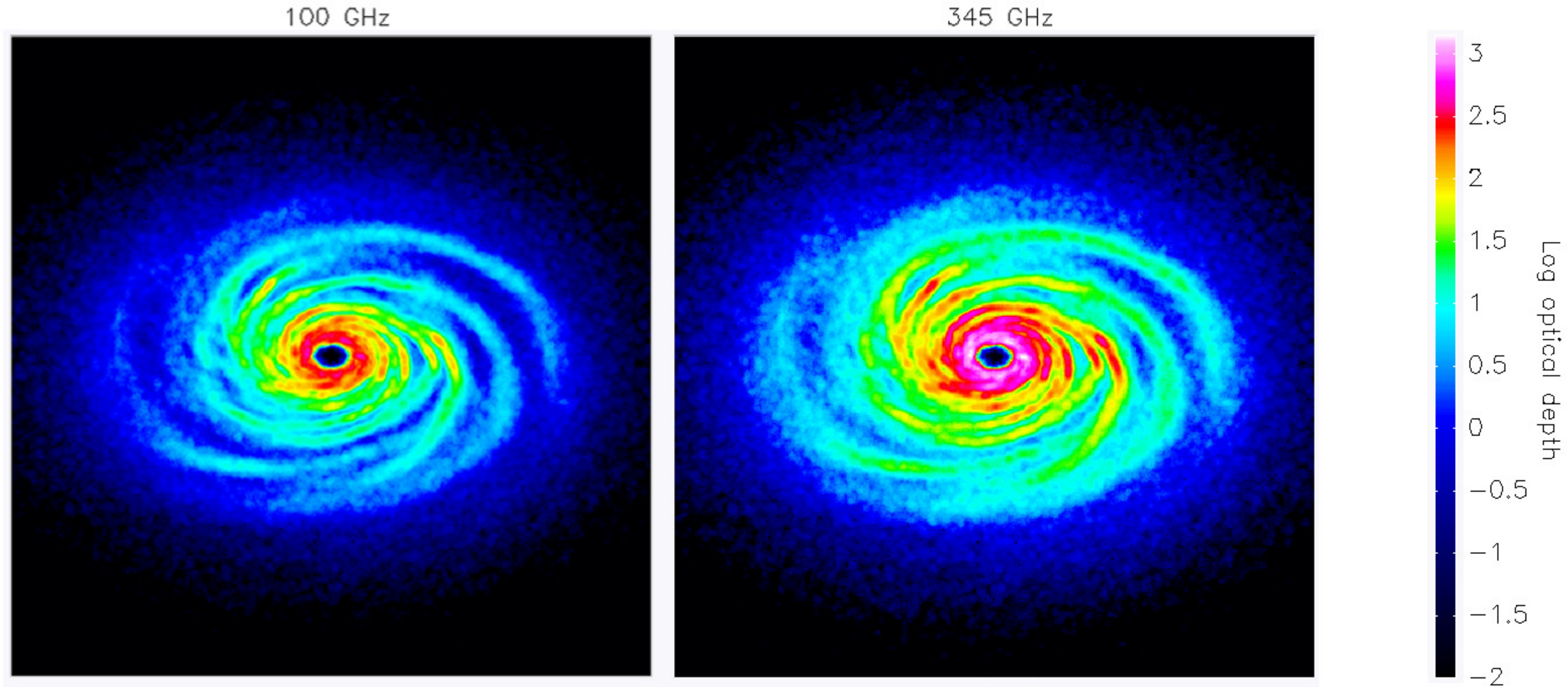}
\caption{Comparison of the optical depth across the disc face at frequencies of 100 GHz (right) and 345 GHz (right) for the disc of radius 100 au with mass equal to $0.25\, \rm{M_{\odot}}$ and a disc-to-star mass ratio $q=0.25$.}
\label{img:opt_depth}
\end{minipage}
\end{figure*}
By analysing the ALMA images at different frequencies, it can be noted that the relative intensity between arm and interarm regions varies with frequency. 
In this respect, there are essentially two important effects to consider in order to specify the trend of the contrast between arm and interarm regions with frequency: the optical depth and the disc temperature. To illustrate the effects of varying the frequency, in Fig. \ref{img:opt_depth} is shown the optical depth at frequencies of 100 and 345 GHz for the disc under consideration.
It can be noted that, while the dense regions corresponding to spiral arms are generally optically thick, the interarm regions transition from being optically thin at frequencies below $\sim$ 345 GHz, to optically thick above 345 GHz. Thus, for $\nu \aplt 345$ GHz, the contrast between arm and interarm regions tends to decrease with increasing frequency.
On the contrary, at higher frequencies ($\apgt 345$ GHz), the contrast increases (see Fig. \ref{img:alma_0.25_r4_m1_incl_45}) because of a temperature effect. In the interarm regions the temperature reaches values low enough to ensure that the Planck law is in the Wien limit for frequencies greater than $345$ GHz. For this reason, the contrast between arm and interarm regions increases with frequency in the range $345$ - $870$ GHz which makes it easier to detect the spiral features. 
This characteristic trend of contrast between arm and interarm regions with frequency can be observed also for compact discs (outer radius of 25 au) with a stellar mass of $0.3\, \rm{M_{\odot}}$ and $q=0.25$. Other discs in our sample are characterized by a temperature high enough to ensure that the Planck law is in the Rayleigh-Jeans limit. In those cases,
the variation in intensity with increasing frequency is only due to the different optical regimes in different parts of the disc. 

There are two interesting features that can be seen from the morphology apparent in the ALMA images. First, note that, while the original density structure of the disc is characterized by a relatively large number of arms, only two spiral arms are visible in the images, because the smaller-sized arms have been washed out by the limited resolution of the observations. Care should thus be taken when interpreting observed morphologies: for example, a clean two-armed spiral is often interpreted as the natural effect of a tidal interaction with a companion star or massive planets, while we show here that it can be easily reproduced in the context of gravitational instability, even though the underlying disc structure is more complex than a two-armed spiral.

Additionally, one can also note a clean clump on the right of the star, along the spiral arm. Now, when comparing the location of the clump with the original density structure, one can notice that it actually corresponds to a mild overdensity at the intersection of two arms. Note that the original simulation did not have any clump, and in fact it is a non fragmenting simulation. The detection of an image such as that in Fig. \ref{img:alma_0.25_r4_m1_incl_45} might be easily mistaken as evidence of fragmentation, while in fact the clump is an artefact resulting from the observational limitations, that in this case overemphasize a mild overdensity in the disc.

\subsubsection{Changing the disc-to-star mass ratio}
\begin{figure*}
\begin{minipage}{\textwidth}
\centering
\includegraphics[scale=0.225]{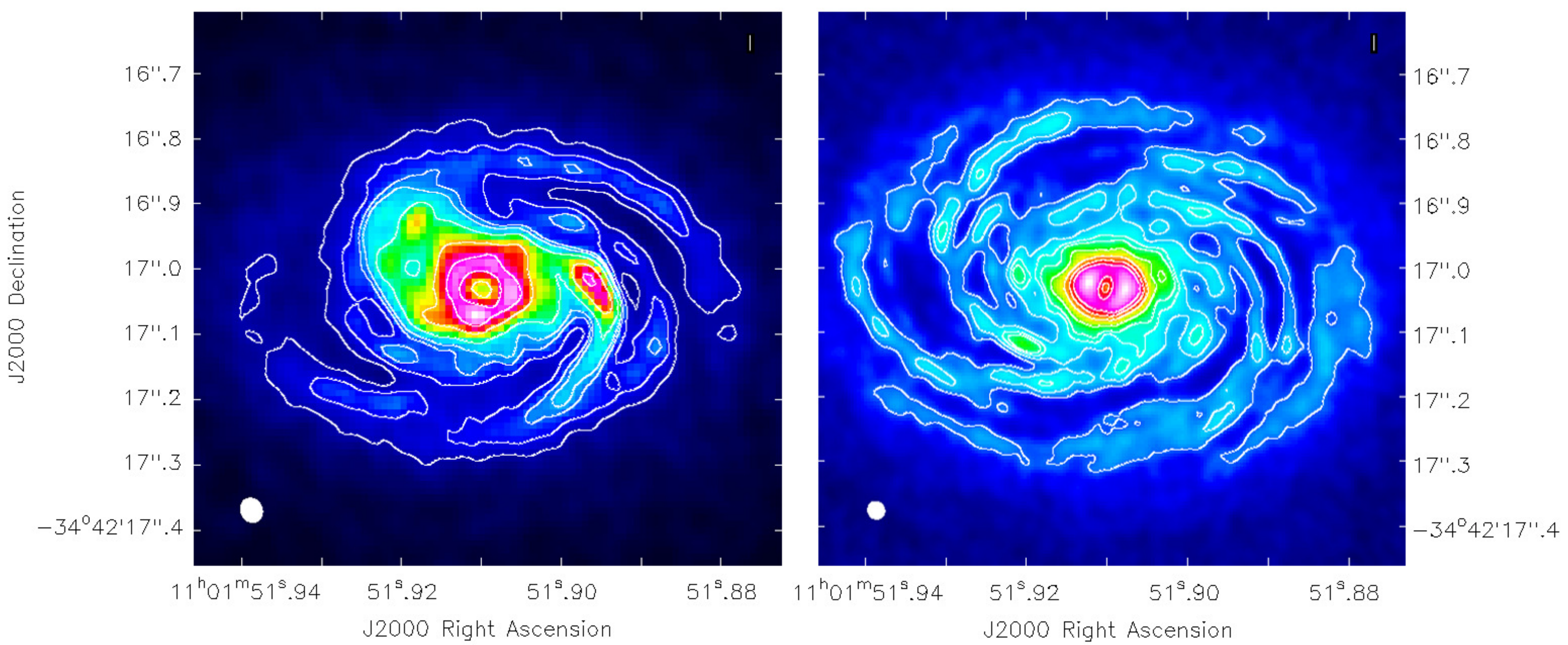}
\caption{ALMA images at 220 GHz of two non face-on (inclination angle of 45\textdegree) compact discs (outer radius of 25 au) located in TW Hydrae around a star of $3\, \rm{M_{\odot}}$ with two different disc-to-star mass ratio: $q=0.25$ (left) and $q=0.1$ (right). Contours are at different multiples ($>4$) of the evaluated rms noise: $38\, \rm{\mu}$Jy/beam (left) and $10\, \rm{\mu}$Jy/beam (right). The white colour in the filled ellipse in the lower left corner indicates the size of the half-power contour of the synthesized beam: $0.033'' \times 0.033''$ (left) and $0.027'' \times 0.025''$ (right).}
\label{img:q_var}
\end{minipage}
\end{figure*}

As previously mentioned (see Section \ref{disc:evol}), discs with different disc-to-star mass ratio are characterized by a different surface density structure. In detail, the number of arms increases and the spiral structure tends to become more close with decreasing disc-to-star mass ratio making it therefore more difficult to detect the spiral features. Thus, in order to test the ALMA capabilities in detecting the spiral structure with different intensity and morphology, we show in Fig. \ref{img:q_var} the ALMA images at 220 GHz of two non face-on (inclination angle of 45\textdegree) compact discs (outer radius of 25 au) located in TW Hydrae around a star of $3\, \rm{M_{\odot}}$ with two different disc-to-star mass ratio: $q=0.25$ (left) and $q=0.1$ (right). As previously mentioned, the temperature profile is closely linked to the mass of the disc (see Fig. \ref{img:temp_prof}). Therefore, for a cooler disc the intensity is lower and we thus need a long observation time to detect spiral features with an appropriate signal-to-noise ratio. Thus, we set the observation time equal to 10 min for the discs with $q=0.25$ and 2 h for the other one (see table \ref{tab:twhya}).
It can be noted that, using an appropriate choice of input parameters, ALMA provides a spatial resolution
sufficient to detect the spiral structure with different arm sizes with an acceptable signal-to-noise ratio, even for a more tightly wound spiral.

\subsubsection{Changing the stellar mass}
\begin{figure*}
\begin{minipage}{\textwidth}
\centering
\includegraphics[scale=0.225]{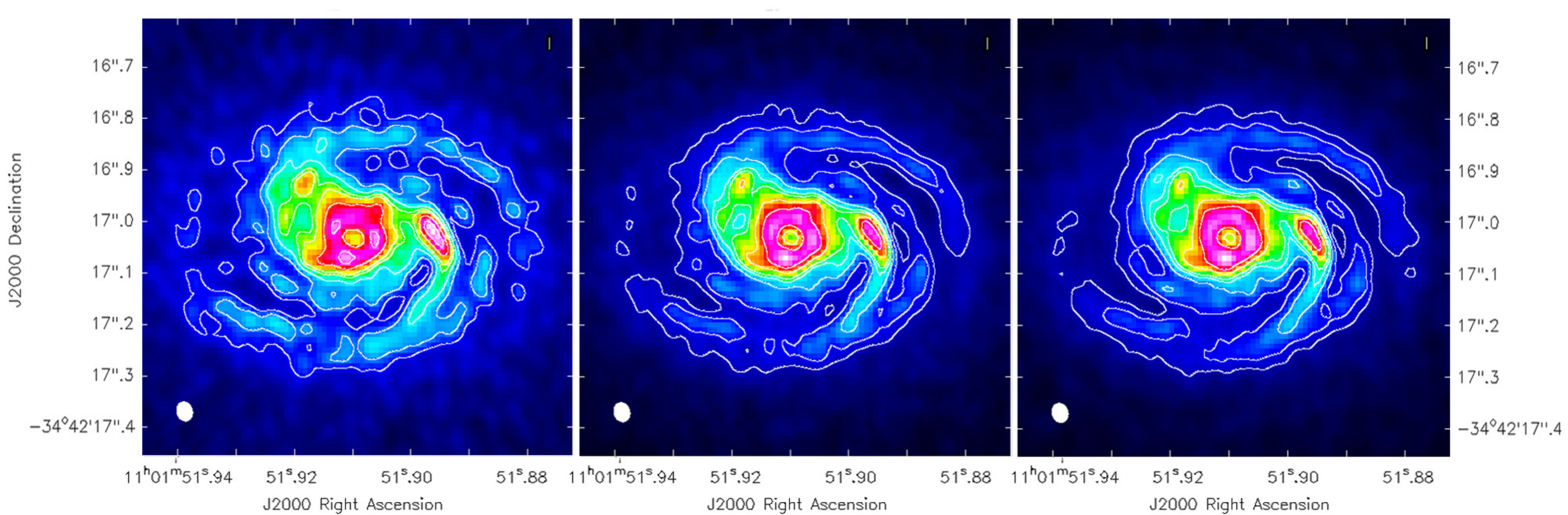}
\caption{ALMA images at 220 GHz for compact discs (outer radius of 25 au) with different star and disc mass located in TW Hydrae. The three images are related to discs with $q=0.25$ with stellar mass of $0.3\, \rm{M_{\odot}}$ (left), $1\, \rm{M_{\odot}}$ (centre) and $3\, \rm{M_{\odot}}$ (right). Contours are at different multiples ($>4$) of the evaluated rms noise: $21\, \rm{\mu}$Jy/beam (left), $40\, \rm{\mu}$Jy/beam (centre) and $38\, \rm{\mu}$Jy/beam (right). The white colour in the filled ellipse in the lower left corner indicates the size of the half-power contour of the synthesized beam: $0.033'' \times 0.033''$.}
\label{img:alma_txtot}
\end{minipage}
\end{figure*}

Here, we show ALMA images for different values of stellar mass for a given value of the disc-to-star mass ratio. 
In Fig. \ref{img:alma_txtot} we show the predicted observability of all the compact discs (outer radius of 25 au) in our sample with $q=0.25$ at the frequency of 220 GHz and located in TW Hydrae star-forming region. 
The three panels are related to non face-on discs (inclination of 45\textdegree) with stellar mass of $0.3\, \rm{M_{\odot}}$ (left), $1\, \rm{M_{\odot}}$ (centre) and $3\, \rm{M_{\odot}}$ (right).
Taking into account the required sensitivity for detection, the ALMA observations are performed with the same choice of antenna configuration and a transit duration of 30 min for the disc with stellar mass $0.3\, \rm{M_{\odot}}$, while for more massive star, due to the greater intensity, the observation time is fixed to 10 min. Moreover, comparing the total resulting flux of discs with 1 and 3 $\mathrm{M_{\odot}}$ (see table \ref{tab:twhya}), note that, as expected, the flux increases with the disc (stellar) mass.
In addition, it can be noted that, due to the lower emission of the less massive disc (left-hand panel in Fig. \ref{img:alma_txtot}), the signal-to-noise ratio on the spiral arms is lower with respect to other cases.    
However, all the simulations demonstrate that resolution and sensitivity of ALMA are amply sufficient to spatially resolve and detect the substructures of the simulated discs with an appropriate choice of observation parameters. 
\subsubsection{Changing the disc inclination}

\begin{figure*}
\begin{minipage}{\textwidth}
\centering
\includegraphics[scale=0.225]{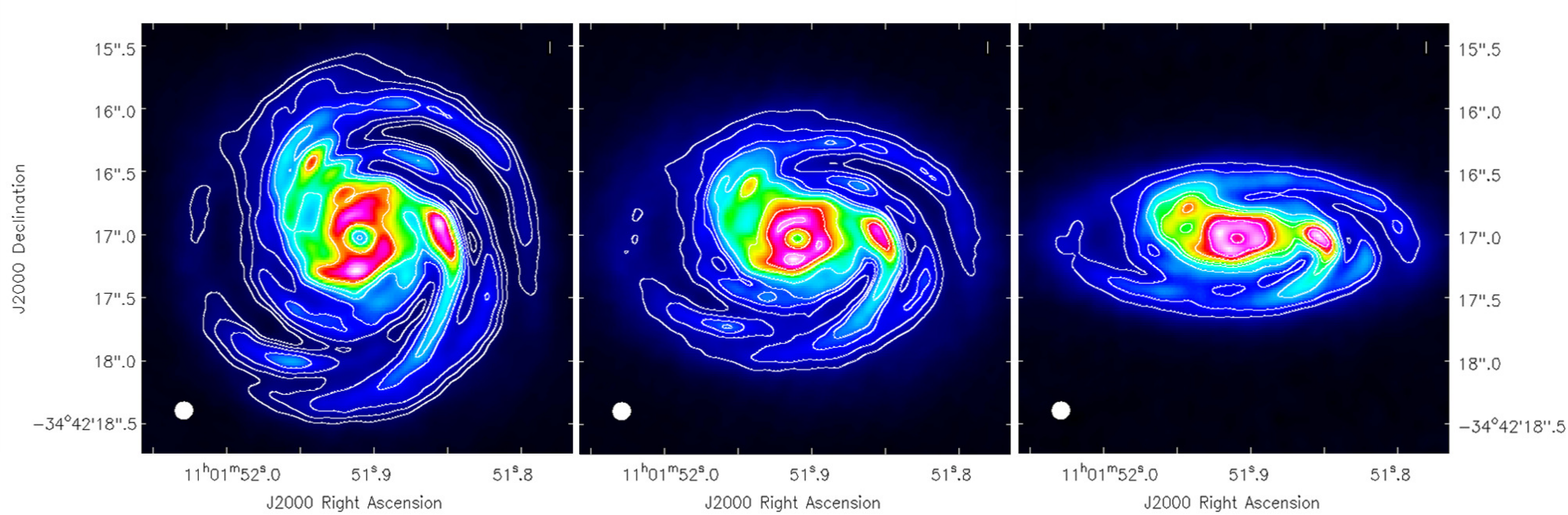}
\caption{ALMA images at 220 GHz for larger discs (outer radius of 100 au) and central mass star of $1\, \rm{M_{\odot}}$ with $q=0.25$ and  with three values of inclination angle: 0\textdegree (left), 45\textdegree (centre) and 65\textdegree (right). Contours are at different multiples ($>4$) of the evaluated rms noise: $\sim 27\, \rm{\mu}$Jy/beam. The white colour in the filled ellipse in the lower left corner indicates the size of the half-power contour of the synthesized beam: $0.147'' \times 0.135''$.}
\label{img:inclination}
\end{minipage}
\end{figure*}

In addition, for what concerns the inclination angle of circumstellar discs, it is expected that the spiral features of the disc becomes less prominent with increasing angle of inclination. This effect makes it more difficult to observe the spiral structures of the accretion disc. We show in Fig. \ref{img:inclination} ALMA images at 220 GHz of the larger disc (outer radius of 100 au) in our sample with a central mass star of $1\, \rm{M_{\odot}}$, $q=0.25$ and with three values of inclination angle: 0\textdegree $\,$(left), 45\textdegree $\,$(centre) and 65\textdegree $\,$(right).
It can be noted that, using the same antenna configuration, the number of the detected spiral arms decreases with inclination angle. As expected, the contrast between arm and interarm regions decreases with increasing angle. In addition, we have also tested the limit of ALMA capability to spatially resolve the spiral pattern. We find that for discs with inclination angle > 75\textdegree, the detected separation between arm and interarm regions vanishes. Assuming that protostellar discs are characterized by a uniform distribution in inclination angle, we conclude that the detection of spiral structure is acceptable for $\sim 80$ per cent of discs.
\begin{table*}
\setlength{\tabcolsep}{5.4pt}
\begin{minipage}{\textwidth}
\begin{center}
\begin{tabular}{ccccccccccccccccc}
\midrule
\midrule
\multicolumn{4}{c} {\footnotesize{Disc model parameters}} &  \multicolumn{6}{c} {\footnotesize{ALMA configurations}} & \footnotesize Duration&\multicolumn{6}{c} {\footnotesize{Total fluxes (mJy)}} \\
\midrule
\,\,\,\footnotesize{q}     &\, \footnotesize{$M_{\star} \,(\rm{M_{\odot}})$}     & \footnotesize{$R_{\rm{out}}$ (au)}  & \footnotesize{Incl. (\textdegree)}  &  \footnotesize45&\footnotesize100&\footnotesize220&\footnotesize345&\footnotesize680&\footnotesize870   &\footnotesize (min)    &  \footnotesize45&\footnotesize100&\footnotesize220&\footnotesize345&\footnotesize680&\footnotesize870 \\ 
\,\,\,\footnotesize{(1)}     &\, \footnotesize{(2)}     & \footnotesize{(3)}  & \footnotesize{(4)}  &  \footnotesize{(5)}&\footnotesize{(6)}&\footnotesize{(7)}&\footnotesize{(8)}&\footnotesize{(9)}&\footnotesize{(10)}   &\footnotesize{(11)}    &  \footnotesize{(12)}&\footnotesize{(13)}&\footnotesize{(14)}&\footnotesize{(15)}&\footnotesize{(16)}&\footnotesize{(17)} \\     
\midrule
\multirow{2}{0.2cm}{\footnotesize0.25}& \multirow{2}{1.2cm}{\,\,\,\,\,\,\,\,\,\,\,\footnotesize1}& \multirow{2}{1.3cm}{\,\,\,\,\,\,\,\,\,\footnotesize25} & \footnotesize0 & \multirow{2}{0.1cm}{\footnotesize28}\,\, &\multirow{2}{0.1cm}{\footnotesize28}  \,\,& \multirow{2}{0.1cm}{\footnotesize28} \,\,  &\multirow{2}{0.1cm}{\footnotesize26} \,\, &\multirow{2}{0.1cm}{\footnotesize22}\,\,  &\multirow{2}{0.1cm}{\footnotesize21} \,\,&\multirow{2}{0.1cm}{\footnotesize30}\,\,& \footnotesize3 & \footnotesize18 &\footnotesize 85&\footnotesize 203 &\footnotesize 708 &\footnotesize 1087 \\
& & & \footnotesize45  & & &    & & &&& \footnotesize2 & \footnotesize10 &\footnotesize 46&\footnotesize 112 &\footnotesize 373 &\footnotesize 571 \\
\midrule
\multirow{2}{0.2cm}{\footnotesize0.25}& \multirow{2}{1.2cm}{\,\,\,\,\,\,\,\,\,\,\,\footnotesize1}& \multirow{2}{1.3cm}{\,\,\,\,\,\,\,\,\footnotesize100} & \footnotesize0 & \multirow{2}{0.1cm}{\footnotesize28} \,&\multirow{2}{0.1cm}{\footnotesize26} \,\, & \multirow{2}{0.1cm}{\footnotesize21} \,\, &\multirow{2}{0.1cm}{\footnotesize20}\,\,\, &\multirow{2}{0.1cm}{\footnotesize15} \,&\multirow{2}{0.1cm}{\footnotesize13}\,\,\,& \multirow{2}{0.1cm}{\footnotesize60}\,\,&\footnotesize6 & \footnotesize46 &\footnotesize 209&\footnotesize 394 &\footnotesize 787 &\footnotesize 1225 \\
& & & \footnotesize45  & & &    & & &&& \footnotesize4 & \footnotesize25 &\footnotesize 120&\footnotesize 212 &\footnotesize 398 &\footnotesize 645 \\
\midrule
\midrule
\end{tabular}
\caption{Disc model, observation parameters and total resulting fluxes for the simulated observations of discs located in Taurus - Auriga star-forming region. See table \ref{tab:twhya} for details.}
\label{tab:taurus}
\end{center}

\end{minipage}
\end{table*}
\begin{table*}
\setlength{\tabcolsep}{5.4pt}
\begin{minipage}{\textwidth}
\begin{center}
\begin{tabular}{ccccccccccccccccc}
\midrule
\midrule
\multicolumn{4}{c} {\footnotesize{Disc model parameters}} &  \multicolumn{6}{c} {\footnotesize{ALMA configurations}} & \footnotesize Duration&\multicolumn{6}{c} {\footnotesize{Total fluxes (mJy)}} \\
\midrule
\,\,\,\footnotesize{q}     &\, \footnotesize{$M_{\star} \,(\rm{M_{\odot}})$}     & \footnotesize{$R_{\rm{out}}$ (au)}  & \footnotesize{Incl. (\textdegree)}  &  \footnotesize45&\footnotesize100&\footnotesize220&\footnotesize345&\footnotesize680&\footnotesize870   &\footnotesize (min)    &  \footnotesize45&\footnotesize100&\footnotesize220&\footnotesize345&\footnotesize680&\footnotesize870 \\    
\,\,\,\footnotesize{(1)}     &\, \footnotesize{(2)}     & \footnotesize{(3)}  & \footnotesize{(4)}  &  \footnotesize{(5)}&\footnotesize{(6)}&\footnotesize{(7)}&\footnotesize{(8)}&\footnotesize{(9)}&\footnotesize{(10)}   &\footnotesize{(11)}    &  \footnotesize{(12)}&\footnotesize{(13)}&\footnotesize{(14)}&\footnotesize{(15)}&\footnotesize{(16)}&\footnotesize{(17)} \\  
\midrule
\multirow{2}{0.2cm}{\footnotesize0.25}& \multirow{2}{1.2cm}{\,\,\,\,\,\,\,\,\,\,\,\footnotesize1}& \multirow{2}{1.3cm}{\,\,\,\,\,\,\,\,\,\footnotesize25} & \footnotesize0 & \multirow{2}{0.1cm}{\footnotesize28}\,\, &\multirow{2}{0.1cm}{\footnotesize28}  \,\,& \multirow{2}{0.1cm}{\footnotesize28} \,\,  &\multirow{2}{0.1cm}{\footnotesize26} \,\, &\multirow{2}{0.1cm}{\footnotesize22}\,\,  &\multirow{2}{0.1cm}{\footnotesize21} \,\,&\multirow{2}{0.1cm}{\footnotesize30}\,\,& \footnotesize4 & \footnotesize21 &\footnotesize 112&\footnotesize 232 &\footnotesize 810 &\footnotesize 1121 \\
& & & \footnotesize45  & & &   & & & && \footnotesize2 & \footnotesize12 &\footnotesize 55&\footnotesize 141 &\footnotesize 417 &\footnotesize 594 \\
\midrule
\multirow{2}{0.2cm}{\footnotesize0.25}& \multirow{2}{1.2cm}{\,\,\,\,\,\,\,\,\,\,\,\footnotesize1}& \multirow{2}{1.3cm}{\,\,\,\,\,\,\,\,\footnotesize100} & \footnotesize0 & \multirow{2}{0.1cm}{\footnotesize28} \,&\multirow{2}{0.1cm}{\footnotesize26} \,\, & \multirow{2}{0.1cm}{\footnotesize21} \,\, &\multirow{2}{0.1cm}{\footnotesize20}\,\,\, &\multirow{2}{0.1cm}{\footnotesize15} \,&\multirow{2}{0.1cm}{\footnotesize13}\,\,\,& \multirow{2}{0.1cm}{\footnotesize60}\,\,& \footnotesize7 & \footnotesize55 &\footnotesize 223&\footnotesize 452 &\footnotesize 812 &\footnotesize 1266 \\
& & & \footnotesize45  & & &    && & && \footnotesize4  &\footnotesize 28&\footnotesize 127 & \footnotesize257 &\footnotesize 435 &\footnotesize 673 \\
\midrule
\midrule
\end{tabular}
\caption{Disc model, observation parameters and total resulting fluxes for the simulated observations of discs located in Ophiucus star-forming region. See table \ref{tab:twhya} for details.}
\label{tab:ophiucus}
\end{center}

\end{minipage}
\end{table*}

\subsubsection{Changing the disc location}

\begin{figure*}
\begin{minipage}{\textwidth}
\centering
\includegraphics[scale=0.217]{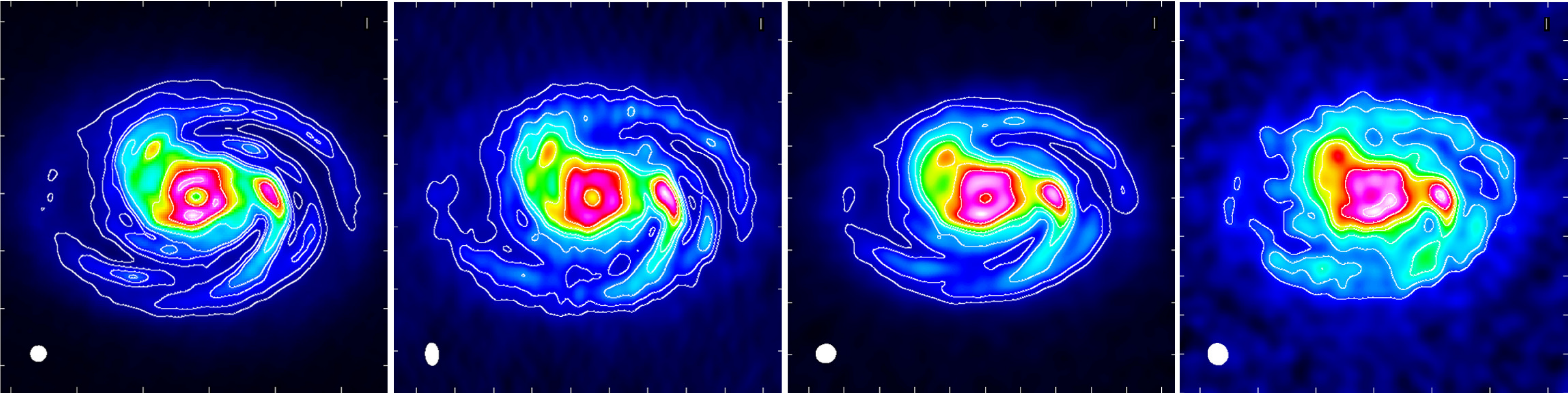}
\caption{ALMA images at 220 GHz for larger discs (outer radius of 100 au) and central mass star of $1 \rm{M_{\odot}}$ with $q=0.25$. From left to right we show observations for discs located in TW Hydrae, Taurus - Auriga, Ophiucus and Orion star-forming region. Contours are at different multiples ($>4$) of the evaluated rms noise: $27$, $20$, $15$ and $10\, \rm{\mu}$Jy/beam. The white colour in the filled ellipse in the lower left corner indicates the size of the half-power contour of the synthesized beam: $0.147'' \times 0.135''$, $0.070'' \times 0.044''$, $0.077'' \times 0.070''$ and $0.025'' \times 0.010''$.}
\label{img:location}
\end{minipage}
\end{figure*}

Finally, in order to create a full 'atlas' of simulated ALMA images of self-gravitating accretion discs, it is necessary to evaluate the detection ability of ALMA for discs located in different star-forming regions.
To this aim, we simulate ALMA observations of a sub-set of discs in our sample located in TW Hydrae, Taurus - Auriga, Ophiucus  and Orion star-forming regions.
In Fig. \ref{img:location} are shown the predicted images at the frequency of 220 GHz of all the larger discs (outer radius of 100 au) in our sample with a stellar mass of $1 \rm{M_{\odot}}$, $q=0.25$ with an inclination angle of 45\textdegree $\,$located in the four previously mentioned star-forming regions. 

The first panel in Fig. \ref{img:location} refers to the case of the TW  Hydrae star-forming region. Due to its closeness to the Solar system ($\sim 50 \,$pc), the resolution capabilities of ALMA are largely sufficient to spatially resolve the substructures for all the discs in the sample over the whole range of observing wavelengths. 

The second and the third panel are related to discs in Taurus - Auriga and Ophiucus ($\sim 140 \,$pc) star-forming regions.  In these cases, the ALMA observations are performed with a transit duration of one hour.
Since these regions are located at nearly the same distance, it is instructive to compare the image reconstruction capabilities of ALMA for disc located in different sky regions with the same choice of observation parameters (see tables \ref{tab:taurus} and \ref{tab:ophiucus}). 
For what concerns the case of discs located in Taurus - Auriga,  it can be noted that the sky position of this source (declination: $\delta=+30$\textdegree) during all the observation duration, produces two important effects involving atmospheric noise and spatial resolution. 
First, the observation of discs located in this star-forming region is more corrupted by atmospheric noise than other cases. This occurs because of the greater atmospheric emission at lower elevations. 
As a consequence, the evaluated rms noise of the ALMA images is greater than what obtained when the source is at lower declination observed with the same input parameter (see the third panel in Fig. \ref{img:location}).
In addition, the atmospheric noise is enhanced especially for higher frequency where the atmospheric transmission plays a crucial role for the quality of observations. As can be seen by comparing the total resulting fluxes for the same discs located in the two considered star-forming regions (see tables \ref{tab:taurus} and \ref{tab:ophiucus}), the effect of the atmospheric noise is essentially a decrease of the detected signal.
The second consequence of the particular source position is related to the spatial resolution provided by an antenna ALMA configurations. The angular resolution of an interferometer is determined by the length of the baselines projected on to a plane normal to the direction of the source. Given that the source elevation is low for all the transit duration, the spatial frequency plane is characterized by an irregular coverage.
For this reason, the synthesized beam is characterized by an half-power contour with two very different values of minor and major axes.
As a consequence, the spatial resolution sufficient to resolve spiral features can be reached only in one direction.
Comparing the second and the third panels of Fig. \ref{img:location}, it is clearly visible that ALMA provides a better reconstruction of image in terms of spatial resolution and signal-to-noise ratio for the disc located in Ophiucus (declination: $\delta=-24$\textdegree). It is worth remarking that the ALMA images for the discs in the two different regions have been carried out using the same antenna configurations and the same total observation durations. Furthermore, comparing the rms noise at each frequency for the same discs located in the two previously mentioned star-forming regions, we find, as expected, a higher rms for the lower elevation sources, especially for intermediate and higher frequencies. 

Finally, the fourth panel of Fig. \ref{img:location} refers to discs in the Orion star-forming region ($\sim 400 \,$pc). We find that the spiral structure for larger discs is detectable at frequencies greater than 100 GHz with an acceptable signal-to-noise ratio. 
In addition, ALMA observation of the disc of radius 25 au (not presented here) show that the spiral structure is not well resolved at low and intermediate frequencies ($45$ - $345$ GHz), whereas at higher frequencies, as noted above, the simulation are probably over-optimistic.  Furthermore, the antenna configuration used to simulate ALMA observation for frequencies lower than 680 GHz are characterized by the longest baseline available. As a consequence, calibration uncertainties and residual phase noise play a crucial role in a realistic image reconstruction. Therefore, the spiral structure on compact discs in the sample is not conclusively detectable using ALMA.

\section{CONCLUSION}
\label{sec:discuss}

In this paper we have shown simulated ALMA observations of a variety of non face-on self-gravitating circumstellar discs models with different properties in mass and radial extension with the aim to demonstrate that the peculiar spiral structure should be detectable using ALMA.  

We conclude that, using a careful choice of observation parameters such as antenna configuration and observation duration, for discs at distances comparable to TW Hydrae ($\sim 50 \,$pc), Taurus - Auriga and Ophiucus ($\sim 140 \,$pc) star-forming regions, the spiral structure is readily detectable by ALMA over a wide range of wavelengths.
However, for discs located in Orion complex ($\sim 400 \,$pc), only the largest discs in the parameter ranges we have explored could be spatially resolved while the smaller ones are characterized by a spiral structure that should not be detectable with ALMA. 
Our simulations also confirm the well-known fact that great care should be taken when interpreting morphological features, such as clumps, in interferometric observations of complex sources. Indeed, in our simulated
observations, an apparent, well-detectable clump is only an artefact
due to finite resolution effects, where the underlying disc structure
only has a mild and transient density enhancement due to the
intersection of two spiral arms.

Spiral and non-axisymmetric structures in protostellar discs have started to be observed relatively frequently in the past few years. In particular, several Class 0 sources, that are more likely to be subject to gravitational instabilities (\citealt{Eisner}, \citealt{Rodriguez2005}, \citealt{Graves}), have been detected to show non-axisymmetric features. 
For example, \citet{Rodriguez2005} have presented mm observations of IRAS 16293-2422B, a Class 0 object located in the Ophiuchus molecular complex, revealing that the gravitational instability might be the origin of the non axisymmetric structures in the outer disc region.
These paper, however, did not fully separate the disc and the envelope contribution to the emission. 
Recently, \citet{Miotello} have reported mm observations of Class 0/I sources where the disc and envelope contribution could be identified. In some cases, the resulting disc appeared to be massive enough to be gravitationally unstable. We should note, however, that the envelope contribution has not been included in our modelling. 

Additionally, a spiral structure has also been observed recently in transitional discs by imaging the distribution of scattered light at near-infrared wavelengths (\citealt{GradyMuto2012}, \citealt{Garufi} and\citealt{GradyMuto2013}) and by high-resolution ALMA observations of molecular line emission (\citealt{Fukagawa}, \citealt{Christiaens2014}). For such evolved discs, one might not expect the disc to be massive enough to be self-gravitating, and thus the more common explanation for the origin of the spiral is the dynamical interaction with a third body, like a companion star or an embedded planet. However, note that  \citet{Fukagawa} and \citet{Christiaens2014} have shown that the outer regions of the transitional disc of the Herbig Fe star HD 142527 is very close to the instability regime, which might indicate that the self-gravitating phase might in some cases last well into the Class II phase.

Moreover, the simplicistic cooling prescription used in the SPH simulations does not take into account the irradiation emitted by the central star. The effect of including irradiation is essentially a decrease in the amplitude of the spiral structure (and thus of the arm-interarm density contrast), which means that our simulations can be somewhat optimistic for strongly irradiated discs.

Furthermore, one of the assumptions of the emission model states that the gas and dust component of the disc are perfectly mixed within the disc. In addition, the dust opacities have been computed using the same level of the grain growth in the whole disc. In this respect, it is thought that the spiral structure induced by the development of gravitational instabilities affects both the radial migration and the grain growth of the dust grains. In detail, since the gas density presents local maxima corresponding to the spiral arms, the radial motion of the dust particles is affected (\citealt{RiceLodato2004}, \citealt{Pinilla}). In addition, the enhanced dust density favours and speeds up collisional growth leading the formation of kilometre-sized planetesimal that can easily collapse and form gravitationally bound larger objects \citep{RiceLodato2006}.
We plan to explore these effects in a forthcoming paper.
\section*{ACKNOWLEDGEMENTS}
We thank Cathie Clarke and Olja Panic for helpful suggestions and for a careful reading of the manuscript. We also thank Eelco van Kampen for support in solving some problems with the ALMA simulator. We would like to thank the referee for useful suggestions. We acknowledge financial support from PRIN MIUR 2010-2011, project ''The Chemical and dynamical evolution of the Milky Way and Local Group Galaxies'', prot. 2010LY5N2T. This work
was partly supported by the ESO Office for Science internship program and by the Italian Ministero
dell'Istruzione, Universit\`a e Ricerca through the grant Progetti Premiali 2012 - iALMA (CUP C52I13000140001). IdG acknowledges support from MICINN (Spain) AYA2011-30228-C03 grant (including FEDER funds).

\bibliography{biblio}
\bibliographystyle{agsm}
\end{document}